\documentclass[reprint,superscriptaddress,showpacs,nofootinbib,amsmath,amssymb]{revtex4-1}
\usepackage{comment}
\usepackage{amssymb}
\usepackage{color}
\usepackage{rotating}
\usepackage{amsmath}
\usepackage{ulem}
\usepackage{overpic}
\usepackage{graphicx}
\usepackage{dcolumn}
\usepackage{bm}
\usepackage{chngpage}
\usepackage{multirow}
\usepackage{slashed}
\usepackage{indentfirst}
\usepackage{booktabs}
\usepackage{amssymb,amsfonts}
\usepackage{amsmath}
\usepackage{color}
\usepackage[colorlinks,citecolor=blue,linkcolor=blue,hypertex,breaklinks=true]{hyperref}

\begin{document}
\title{Implication of odd-even staggering in the charge radii of calcium isotopes}
\author{Rong An}
\email[Corresponding author:]{rongan@nxu.edu.cn}
\affiliation{School of Physics, Ningxia University, Yinchuan 750021, China}
\affiliation{Guangxi Key Laboratory of Nuclear Physics and Technology, Guangxi Normal University, Guilin, 541004, China}
\affiliation{Key Laboratory of Beam Technology of Ministry of Education, School of Physics and Astronomy, Beijing Normal University, Beijing 100875, China}

\author{Xiang Jiang}
\affiliation{College of Physics and Optoelectronic Engineering, Shenzhen University, Shenzhen 518060, China}

\author{Na Tang}
\affiliation{School of Physics, Ningxia University, Yinchuan 750021, China}
\affiliation{Key Laboratory of Beam Technology of Ministry of Education, School of Physics and Astronomy, Beijing Normal University, Beijing 100875, China}

\author{Li-Gang Cao}
\email[Corresponding author:]{caolg@bnu.edu.cn}
\affiliation{Key Laboratory of Beam Technology of Ministry of Education, School of Physics and Astronomy, Beijing Normal University, Beijing 100875, China}
\affiliation{Key Laboratory of Beam Technology of Ministry of Education, Institute of Radiation Technology, Beijing Academy of Science and Technology, Beijing 100875, China}

\author{Feng-Shou Zhang}
\email[Corresponding author:]{fszhang@bnu.edu.cn}
\affiliation{Key Laboratory of Beam Technology of Ministry of Education, School of Physics and Astronomy, Beijing Normal University, Beijing 100875, China}
\affiliation{Key Laboratory of Beam Technology of Ministry of Education, Institute of Radiation Technology, Beijing Academy of Science and Technology, Beijing 100875, China}
\affiliation{Center of Theoretical Nuclear Physics, National Laboratory of Heavy Ion Accelerator of Lanzhou, Lanzhou 730000, China}

\date{\today}

\begin{abstract}
  Inspired by the profoundly observed odd-even staggering and the inverted parabolic-like shape in charge radii along calcium isotopic chain, the ground state properties of calcium isotopes are investigated by constraining the root-mean-square (rms) charge radii under the covariant energy density functionals with effective forces NL3 and PK1. In this work, the pairing correlations are tackled by solving the state-dependent Bardeen-Cooper-Schrieffer equations. The calculated results suggest that the binding energies obtained by the radius constraint method have been slightly changed by about $0.2\%$. But for charge radii, the corresponding results deriving from NL3 and PK1 forces have been increased by about $1.0\%$ and $2.0\%$, respectively. This means that charge radius is a more sensitive quantity in the calibrated protocol. Meanwhile, it is found that the reproduced charge radii of calcium isotopes are attributed to the rather strong isospin dependence of effective potential. The odd-even oscillation behavior can also be presented in the proton Fermi energies along calcium isotopic family, but keep opposite trends with respect to the corresponding binding energies and charge radii. As encountered in charge radii, the weakened odd-even oscillation behavior is still emerged from the proton Fermi energies at the neutron numbers $N=20$ and $28$ as well, but not in binding energies.
\end{abstract}

%\pacs{25.70.Jj,24.10.-i}

\maketitle

\section{Introduction}\label{sec0}

Accurate description of nuclear charge radii derived from charge density distributions plays an important role in theoretical studies. It is generally mentioned that the fine structure of nuclear size can encode the information about the shape-phase transition~\cite{PhysRevC.27.2282,PhysRevLett.54.1991,LALAZISSIS199635,PhysRevC.82.061302,PhysRevLett.117.172502,Marsh:2018wxs,
PhysRevLett.127.192501,An_2023,MUN2023138298}, halo structure~\cite{PhysRevLett.102.062503,PhysRevLett.101.252502}, and the emergence of magic numbers~\cite{BAGCHI2019251,PhysRevLett.129.142502}, etc.
Moreover, recent studies suggest that charge radii of mirror partner nuclei can provide an alternative access to probe the equation of state (EoS) of isospin asymmetric nuclear matter~\cite{PhysRevC.88.011301,PhysRevLett.119.122502,
PhysRevResearch.2.022035,PhysRevLett.127.182503,PhysRevLett.130.032501,XU2022137333,PhysRevC.107.034319,nuclscitech34.119,
PhysRevC.108.015802,Konig:2023rwe,GAUTAM2024122832}.
With the advent of radioactive beam facilities, more data about charge radii are accumulated over the past decades~\cite{ANGELI201369,LI2021101440}.
Thus this challenges us to recognize the fine structure of atomic nuclei with high-precision charge radii data in nuclear models.

The global features of nuclear charge radii can be generally described by the $A^{1/3}$ or $Z^{1/3}$ law~\cite{A.Bohr,Zhang:2001nt}, with $A$ and $Z$ the mass and proton numbers.
However, these empirical formulae are not valid any more for nuclei with larger neutron to proton ratios.
In addition, it is difficult to inspect the microscopic implications, e.g., the proton and neutron density distributions.
The same scenarios can also be encountered in the Bayesian neural networks~\cite{PhysRevC.105.014308,DONG2023137726}, in which the microscopic aspects cannot be completely captured and the extrapolating ability is restricted due to the limited databases.
Along a long isotopic chain, the evolution of nuclear charge radii has performed regularity and irregularity behaviors~\cite{PhysRevC.100.044310,GarciaRuiz:2019cog}.
Especially in calcium isotopes, the inverted parabolic-like shape and strong odd-even oscillations in charge radii can be apparently observed between $^{40}$Ca and $^{48}$Ca isotopes~\cite{ANGELI201369}.
Furthermore, such peculiar feature of the strong odd-even oscillations in charge radii persists toward the neutron-deficient region~\cite{Miller2019}.
It should be also mentioned that the values of charge radii for $^{40}$Ca (3.4776~fm) and $^{48}$Ca (3.4777~fm) isotopes are almost equivalent, and the shell quenching effect in charge radii has been observed at the neutron number $N=28$ but not at $N=20$.
Across the $^{48}$Ca isotope, the rapid increase of charge radii is performed until to $^{52}$Ca.
The values of charge radii for $^{44}$Ca (3.5188~fm) and $^{50}$Ca (3.5192~fm) are almost identical~\cite{ANGELI201369,LI2021101440}.
For $^{52}$Ca, as a candidate of doubly magic nuclei~\cite{PhysRevC.31.2226,PhysRevC.74.021302,PhysRevLett.109.032506,nature498}, the shell closure effect of the charge radii cannot be expected naturally~\cite{Ruiz2016}.
These particular features can provide insights into the fundamental interactions in theoretical researches.

Plenty of methods have been undertaken to elucidate the global trend of charge radii tentatively along calcium isotopes.
$Ab$ initio calculations with chiral effective field theory interactions, such as NNLOsat~\cite{PhysRevC.91.051301} with SRG1 and SRG2 interactions~\cite{PhysRevC.83.031301}, cannot reproduce the discontinuous changes of the charge radii along calcium isotopes well~\cite{Ruiz2016}.
The major classes of covariant energy density functionals (CEDFs) have made greatly success in describing the ground state properties of finite nuclei over the whole periotic table~\cite{Geng:2003pk,XIA20181,ZHANG2022101488,PhysRevC.104.064313,ADNDT.158.101661,PhysRevC.104.054312,ADNDT.156.101635}.
However, the description and explanation for some fine structures, such as the odd-even staggering (OES) of charge radii in calcium isotopes, are still not adequate yet.
As shown in Ref.~\cite{PhysRevC.53.1599}, the effect of short-rang correlations should be considered properly in describing the charge radii and the depletion of the nuclear Fermi surface.
Miller, $et~al.$ further point out that neutron-proton short-range correlations can have an influence on determining the nuclear charge radius~\cite{Miller:2018mfb}.
The short-range correlation between neutrons and protons implies the fluctuation of the fractional occupation probabilities for the states around Fermi surface~\cite{PhysRevC.101.065202}.
This suggests that the appropriate isospin-dependence interactions should be taken into account properly in describing the nuclear size~\cite{COSYN2021136526}.

The neutron-proton correlations derived from the Casten factor can reproduce the shell quenching phenomena in nuclear charge radii as well~\cite{PhysRevLett.58.658,Angeli_1991,Sheng:2015poa}. In this method, the isospin-dependence interactions deduced from the valence neutrons and protons can actually improve the predicted validity of nuclear charge radii in the calibrated protocol~\cite{PhysRevC.105.014308,DONG2023137726,Xian:2024rmr}.
The similar approach can also be shown in Ref.~\cite{PhysRevC.102.024307} where the correlation between the neutron and proton pairs around Fermi surface has been added into the root-mean-square (rms) charge radii formula.
This modified expression can reproduce the trend of changes of nuclear charge radii and OES well~\cite{PhysRevC.105.014325,An:2023kni}.
In the Fayans energy density functional model, the discontinuous behavior of charge radii in calcium isotopes can be described well~\cite{Miller2019}. These local variations are attributed to the additional gradient terms in its pairing part~\cite{PhysRevC.95.064328}.
Shown in Ref.~\cite{CAURIER2001240} that the abnormal changes of charge radii in the calcium isotopes can be described well in the shell model calculations by considering the proton excitation from the $sd$ shell to the $pf$ shell.
Recent study has analyzed the radial and orbital contributions in reproducing the charge radii of calcium isotopes~\cite{Inakura:2024mri}. Meanwhile, it points out that charge radii in the whole region $^{36-48}$Ca cannot be reproduced perfectly through energy density functionals (EDFs).
In order to inspect the anomalous behaviors in charge radii of calcium isotopes, further research should be performed at the relativistic mean field level.

In this work, we focus on the underlying mechanisms of such enhancement OES and the inverted parabolic-like shape in charge radii of calcium isotopes through mean-field framework.
To facilitate the underlying mechanism of OES in nuclear size, the constraint calculations on the root-mean-square (rms) charge radii are performed along calcium isotopes.
The binding energies and two-neutron separation energies obtained by the radius constraint method are also examined.
Moreover, the rather strong isospin dependence of effective potential is also shown. We further review the evolution of the neutron and proton Fermi energies with the increasing neutron numbers.
The inverse odd-even oscillations of proton Fermi energies are shown against the OES in the charge radii of calcium isotopes.

%The neutron Fermi energies along a long isotopic chain are almost not changed with considering the radius constraint calculations,
%but inverse OES is found dramatically in the neutron-skin thickness.

The structure of the paper is organized as the follows.
Sec~2~is devoted to presenting the theoretical framework briefly.
In Sec~3, the numerical results and discussion are provided.
Finally, a summary and outlook is given in Sec~4.
\section{Theoretical framework}\label{sec1}

The covariant energy density functionals (CEDFs) have been widely used to investigate various physical phenomena, such as halo structure~\cite{PhysRevC.68.034323,PhysRevC.82.011301,SUN2018530,PhysRevC.107.L041303,ZHANG2023138112}, pseudospin symmetry~\cite{LIANG20151,PhysRevC.96.054306,PhysRevC.100.051301}, hypernuclei~\cite{RONG2020135533},
resonant states~\cite{zhang2007,Cao:2003yn}, collective excited states~\cite{PhysRevC.67.034312,PhysRevC.69.054303,PhysRevC.105.034330,PhysRevLett.131.202502}, location of dripline~\cite{An:2020wcv,PhysRevC.104.024331,PhysRevC.108.054305}, nuclear fission~\cite{li2023covariant,zhang2023ternary}, low-lying states~\cite{Wang_2022}, and neutron drops~\cite{PhysRevC.94.041302}, nuclear weak decay~\cite{SciBull.69.2017,PhysLettB.855.138796}, etc. In this work, the non-linear self-coupling Lagrangian density is employed~\cite{Ring:1997tc},
where the constituent nucleons are described as Dirac particles which interact via the exchange of $\sigma$, $\omega$ and $\rho$ mesons. The electromagnetic field is ruled by the exchange of photons. The effective Lagrangian density has been recalled as follows,
\begin{small}
\begin{eqnarray}\label{lag1}
\mathcal{L}&=&\bar{\psi}[i\gamma^\mu\partial_\mu-M-g_\sigma\sigma
-\gamma^\mu(g_\omega\omega_\mu+g_\rho\vec
{\tau}\cdotp\vec{\rho}_{\mu}+e\frac{1-\tau_{3}}{2}\bm{A}_\mu)]\psi\nonumber\\
&&+\frac{1}{2}\partial^\mu\sigma\partial_\mu\sigma-\frac{1}{2}m_\sigma^2\sigma^2
-\frac{1}{3}g_{2}\sigma^{3}-\frac{1}{4}g_{3}\sigma^{4}\nonumber\\
&&-\frac{1}{4}{\bm{\Omega}}^{\mu\nu}{\bm{\Omega}}_{\mu\nu}+\frac{1}{2}m_{\omega}^2\omega_\mu\omega^\mu+\frac{1}{4}c_{3}(\omega^{\mu}\omega_{\mu})^{2}\nonumber\\
&&-\frac{1}{4}\vec{\bm{R}}_{\mu\nu}\cdotp\vec{\bm{R}}^{\mu\nu}+\frac{1}{2}m_\rho^2\vec{\rho}^\mu\cdotp\vec{\rho}_\mu
+\frac{1}{4}d_{3}(\vec{\rho}^{\mu}\vec{\rho}_{\mu})^{2}\nonumber\\
&&-\frac{1}{4}\bm{F}^{\mu\nu}\bm{F}_{\mu\nu},
\end{eqnarray}
\end{small}
where $M$ is the mass of nucleon and $m_{\sigma}$, $m_{\omega}$, and $m_{\rho}$, are the masses of the $\sigma$, $\omega$ and $\rho$ mesons, respectively. Here, $g_{\sigma}$, $g_{\omega}$, $g_{\rho}$, $g_{2}$, $g_{3}$, $c_{3}$ and $d_{3}$ are the coupling constants for $\sigma$, $\omega$ and $\rho$ mesons, respectively, while $e^{2}/4\pi=1/137$ is the fine structure constant.
The field tensors for the vector mesons and photon fields are defined as ${\bm\Omega}_{\mu\nu}=\partial_{\mu}\omega_{\nu}-\partial_{\nu}\omega_{\mu}$,
$\vec{\bm{R}}_{\mu\nu}=\partial_{\mu}\vec{\rho}_{\nu}-\partial_{\nu}\vec{\rho}_{\mu}-g_{\rho}(\vec{\rho}_{\mu}\times\vec{\rho}_{\nu})$ and $\bm{F}_{\mu\nu}=\partial_{\mu}\bm{A}_{\nu}-\partial_{\nu}\bm{A}_{\mu}$.

The Hamiltonian derived from the variational principle can be expressed as follows,
\begin{eqnarray}
{\hat{\mathcal{H}}}=\bm{\alpha}\cdot\bm{p}+V(\bm{r})+\beta{[M+S(\bm{r})]},
\end{eqnarray}
where $S(\bm{r})$ and $V(\bm{r})$ represent the effective fields~\cite{Ring:1997tc}.
The bulk properties of finite nuclei are calculated by the effective forces NL3~\cite{PhysRevC.55.540} and PK1~\cite{PhysRevC.69.034319}.

As mentioned above, the discontinuous behavior of charge radii along calcium isotopes cannot be reproduced well by the relativistic mean field calculations.
To further explore the microscopic mechanism in reproducing the abnormal behaviors in the charge radii of calcium isotopes, a radius constraint calculation is performed in this work.
Therefore, the Hamiltonian can be rewritten as the following expression,
\begin{eqnarray}
\hat{\mathcal{H'}}=\hat{\mathcal{H}}-\lambda\langle{R_{\mathrm{ch}}}\rangle,
\end{eqnarray}
where $\lambda$ is the spring constant.
The quantity of $\langle{R_{\mathrm{ch}}}\rangle$ can be calculated through (in units of fm$^{2}$)
\begin{eqnarray}\label{cp}
R_{\mathrm{ch}}^{2}=\langle{r_{\mathrm{p}}^{2}}\rangle+0.64~\mathrm{fm^{2}}.
\end{eqnarray}
The first term represents the charge density distributions of point-like protons and the second one is that due to the finite size effect of protons~\cite{Gambhir:1989mp}.
The accuracy of the convergence is determined by the self-consistent iteration in binding energy, which
is lower than $10^{-6}$ MeV.

\section{Results and Discussion}\label{sec2}
At the mean field level, the bulk properties of finite nuclei can be described well~\cite{RevModPhys.75.121,VRETENAR2005101}.
However, for charge radii of calcium isotopes, the abnormal behaviors cannot be reproduced well, especially in relativistic mean-field model~\cite{Geng:2003pk,XIA20181,ZHANG2022101488,PhysRevC.104.064313}.
This is attributed to the underlying mechanism which is not clear to comprehend these anomalous behaviors in nuclear charge radii.
In this work, the global trend of nuclear charge radii along calcium isotopic chain is discussed by radius constraint calculation.
The pairing correlations can be treated by solving the state-dependent Bardeen-Cooper-Schrieffer (BCS) equations which can describe the ground state properties of finite nuclei well~\cite{Geng:2003pk}.
The pairing strength is chosen to be $V_{0}=350$ MeV fm$^{3}$ for NL3 force, but $V_{0}=380$ MeV fm$^{3}$ for PK1 parametrization set.
The calculated results show that the quadrupole deformation of calcium isotopes cannot be changed with and without considering radius constraint, namely keep almost spherical shape.

\begin{table*}[htb!]
\caption{Charge radii ($R_{\mathrm{ch}}$) and binding energies (BE) of calcium isotopes obtained by the approaches with and without considering the charge radius constraint calculations are listed for NL3 and PK1 effective interactions, respectively. The experimental data for charge radii~\cite{ANGELI201369,LI2021101440} and binding energy~\cite{cpc2021} are also shown for comparison.}\label{tab1}
 \doublerulesep 0.1pt \tabcolsep 6pt
	\begin{tabular}{rcccccccccc}
		\hline
		\hline
      {} &  \multicolumn{4}{c}{$R_{\mathrm{ch}}$ (fm)}& & \multicolumn{4}{c}{BE (MeV)} & \\
       \cline{2-5} \cline{7-10}
		$N$ & NL3 &  PK1  & NL3 & PK1 &  Exp.  & NL3 &  PK1  & NL3 & PK1 &  Exp. \\
		\hline
       {}  & \multicolumn{2}{c}{Without}  &  \multicolumn{2}{c}{With} & &\multicolumn{2}{c}{Without} &
        \multicolumn{2}{c}{With} & \\
       \cline{2-3} \cline{4-5} \cline{7-8} \cline{9-10}		
        16&3.4596&3.4463&3.4493&3.4493&3.4493&279.44&280.15&279.43&280.15&281.37\\
        17&3.4595&3.4428&3.4480&3.4480&3.4480&294.72&295.05&294.71&295.04&296.13\\
        18&3.4624&3.4430&3.4661&3.4661&3.4661&311.11&312.41&311.11&312.32&313.12\\
        19&3.4650&3.4431&3.4595&3.4595&3.4595&325.96&326.90&325.96&326.79&326.42\\
        20&3.4692&3.4445&3.4776&3.4776&3.4776&341.91&342.79&341.90&342.62&342.05\\
        21&3.4678&3.4438&3.4780&3.4780&3.4780&350.66&351.56&350.56&351.18&350.42\\
        22&3.4663&3.4434&3.5081&3.5081&3.5081&361.32&363.04&361.03&362.33&361.90\\
        23&3.4659&3.4438&3.4954&3.4954&3.4954&369.66&371.12&369.83&370.67&369.83\\
        24&3.4661&3.4454&3.5179&3.5179&3.5179&379.98&381.87&380.09&380.90&380.96\\
        25&3.4667&3.4468&3.4944&3.4944&3.4944&387.96&389.36&388.10&388.91&388.37\\
        26&3.4678&3.4494&3.4953&3.4953&3.4953&397.88&399.40&398.18&398.97&398.77\\
        27&3.4703&3.4522&3.4783&3.4783&3.4783&405.69&406.36&405.68&406.23&406.05\\
        28&3.4711&3.4542&3.4771&3.4771&3.4771&415.07&415.63&415.06&415.53&416.00\\
        29&3.4810&3.4640&3.4917&3.4917&3.4917&419.78&419.91&419.76&419.76&421.15\\
        30&3.4912&3.4766&3.5186&3.5186&3.5186&425.33&425.53&425.34&425.20&427.51\\
        31&3.5023&3.4870&3.5335&3.5335&3.5335&428.77&428.61&428.70&428.21&432.32\\
        32&3.5082&3.4998&3.5531&3.5531&3.5531&435.18&434.66&434.94&434.16&438.33\\
        \hline
		\hline	
	\end{tabular}
\end{table*}
As shown in Table~\ref{tab1}, charge radii and binding energies of calcium isotopes are listed before and after the constrained calculations, respectively. Here, one should be mentioned that binding energies are slightly changed for the results obtained by the NL3 and PK1 effective forces.
For the $^{44}$Ca isotope, the binding energies obtained by radius constraint method have been reduced less than $0.1\%$ for NL3 set and about $0.2\%$ for PK1 set. After making the constrained calculations, the binding energies for $^{40-48}$Ca isotopes are in closer agreement with the experimental data.
However, the obvious deviations can be encountered in the charge radii. Such as the case in $^{44}$Ca isotope, the charge radius obtained by the NL3 effective force has been increased about $1\%$, but about $2\%$ for the PK1 force. This difference attributes to the fact that the charge radii obtained by the PK1 set are systematically underestimated with respect to the results obtained by NL3 set if the radius constraint approach cannot be employed~\cite{PhysRevC.102.024307}. Especially the effective force NL3 can almost reproduce the charge radii of $^{40}$Ca (3.4692 fm) and $^{48}$Ca (3.4711 fm) isotopes, but apparent discrepancy occurs for PK1 set, namely 3.4445 fm for $^{40}$Ca and 3.4542 fm for $^{48}$Ca.
Furthermore, these results suggest that charge radii are more sensitive quantities in the calibrated protocol. This seems to provide a guideline to the artificial neural networks in learning experimental data.
Recent studies suggest that the structure of the input data can also influence the predicted quantities in the artificial neural networks~\cite{PhysRevC.106.L021301,nst.33.153}. Meanwhile, it is worthwhile to mention that the nucleon's intrinsic electromagnetic structure plays an indispensable role in reproducing the charge radii of finite nuclei~\cite{PhysRevC.110.064319}.

The odd-even staggering behaviors in charge radii and nuclear masses are generally observed throughout the whole
nuclide chart~\cite{ANGELI201369,LI2021101440,cpc2021}.
To facilitate the local variations of nuclear charge radii and binding energies, the three-point formula is recalled as follows,
\begin{small}
\begin{eqnarray}\label{oef}
\Delta_{\mathcal{Y}}(N,Z)=\frac{1}{2}[2\mathcal{Y}(N,Z)-\mathcal{Y}(N-1,Z)-\mathcal{Y}(N+1,Z)],
\end{eqnarray}
\end{small}
where $\mathcal{Y}(N,Z)$ represents the specific physical quantity of a nucleus with neutron number $N$ and proton number $Z$.

\begin{figure*}[htbp]
\centering
\includegraphics[scale=0.5]{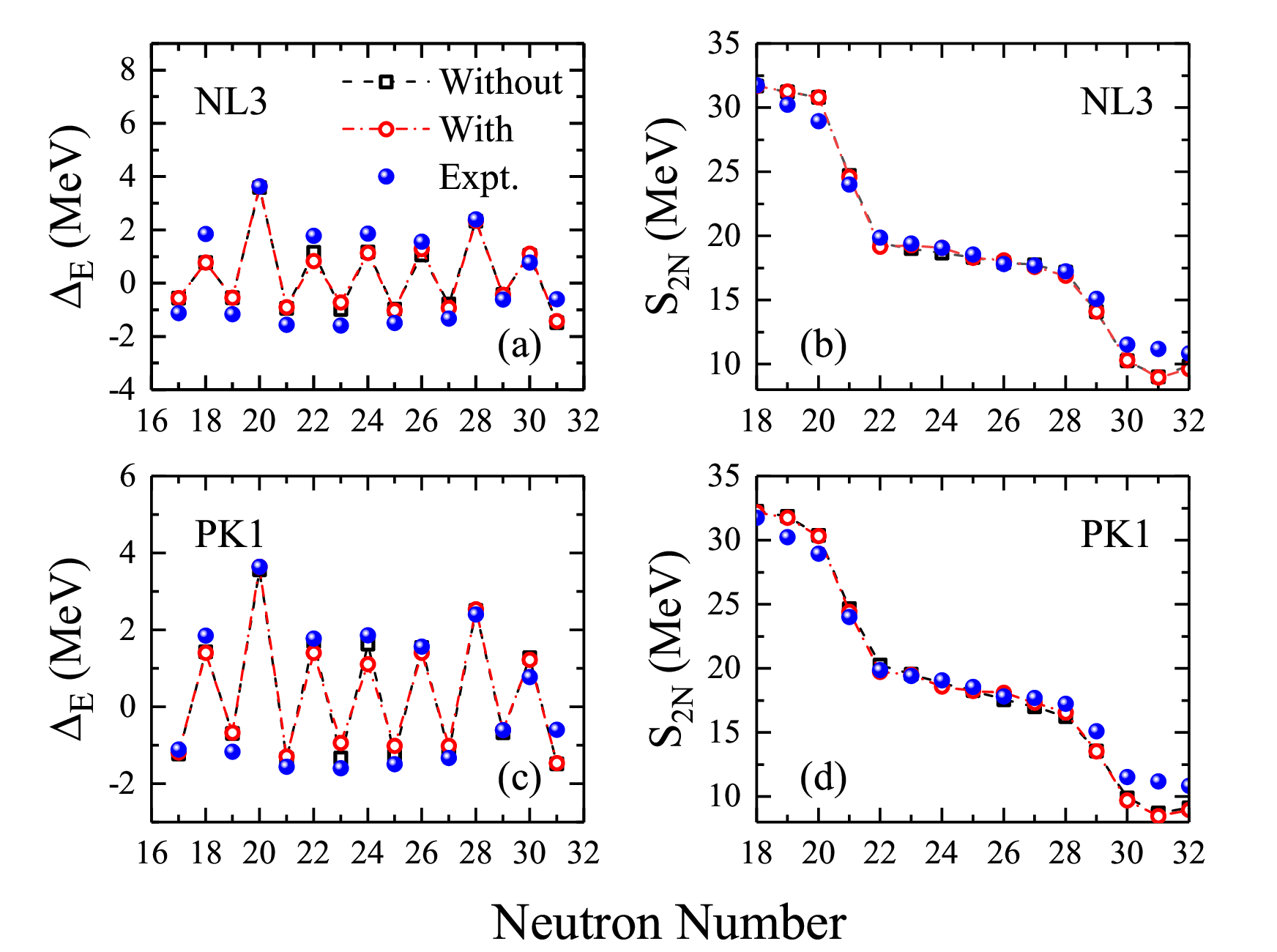}
\caption{(Color online) Odd-even staggering of binding energies and two-neutron separation energies of calcium isotopes as a function of neutron numbers are depicted by both NL3 (a, b) and PK1 (c, d) effective forces. The corresponding experimental data are taken from Ref.~\cite{cpc2021}.} \label{fig1}
\end{figure*}

The OES of binding energies calculated by Eq.~(\ref{oef}) are shown in Fig.~\ref{fig1}~(a) and (c) with the effective forces NL3 and PK1, respectively.
Here one can find that the OES in the evaluated nuclear mass agrees with the experimental data very well.
With and without considering the constraint on rms charge radii, the OES of binding energies obtained by both methods can give almost similar trend.
In addition, one can find that the amplitudes of OES in binding energies are enlarged at the neutron numbers $N=20$ and $N=28$ with respect to the neighboring counterparts.
This emerged signature in the empirical mass gap could be generally used to identify the magic numbers~\cite{cpc2021}.

In Fig.~\ref{fig1}~(b) and (d), two-neutron separation energies of calcium isotopes are plotted as a function of neutron numbers with effective forces NL3 and PK1. The similar trend can be found with both of the effective forces NL3 and PK1, especially the sharp decreases of the two-neutron separation energies can also be reproduced well at the neutron numbers $N=20$ and $N=28$.
As shown in Fig.~\ref{fig1}~(a) and (c), the enlarged OES in binding energies results from the strong shell closure.
However, the odd-even oscillation amplitudes of charge radii are weakened at the $N=20$ and $N=28$ shell closure~\cite{PhysRevC.102.024307,An:2023kni}.
As demonstrated in Ref.~\cite{PhysRevC.105.014325}, this weakening OES behavior in nuclear charge radii can be observed generally at the fully filled shells.
This may provide a signature to feature the magicity of a nucleus from the aspect of nuclear size.
As a candidate of doubly magic nuclei, the unexpectedly larger charge radius has been observed in $^{52}$Ca ~\cite{Ruiz2016}.
Thus the charge radius of $^{53}$Ca isotope is urgently required.
The latest study suggests that the abrupt increase of nuclear charge radii along Sc isotopic chain can be observed notably across neutron number $N=20$~\cite{PhysRevLett.131.102501}, but this phenomenon is absent in the neighboring Ca isotopes.

As demonstrated in Ref.~\cite{PhysRevC.102.024307}, the local variations of charge radii along Ca isotopes cannot be reproduced well under the relativistic mean field model with Eq.~(\ref{cp}). Actually, before the constrained calculations, the results obtained by the NL3 and PK1 effective forces give the underestimated OES amplitudes.
Based on the radius constraint calculations, the calculated results can cover the experimental charge radii.
Thus the OES behavior of charge radii between the neutron numbers $N=20$ and $N=28$ are naturally predicted.

\begin{figure}[htbp]
\centering
\includegraphics[scale=0.6]{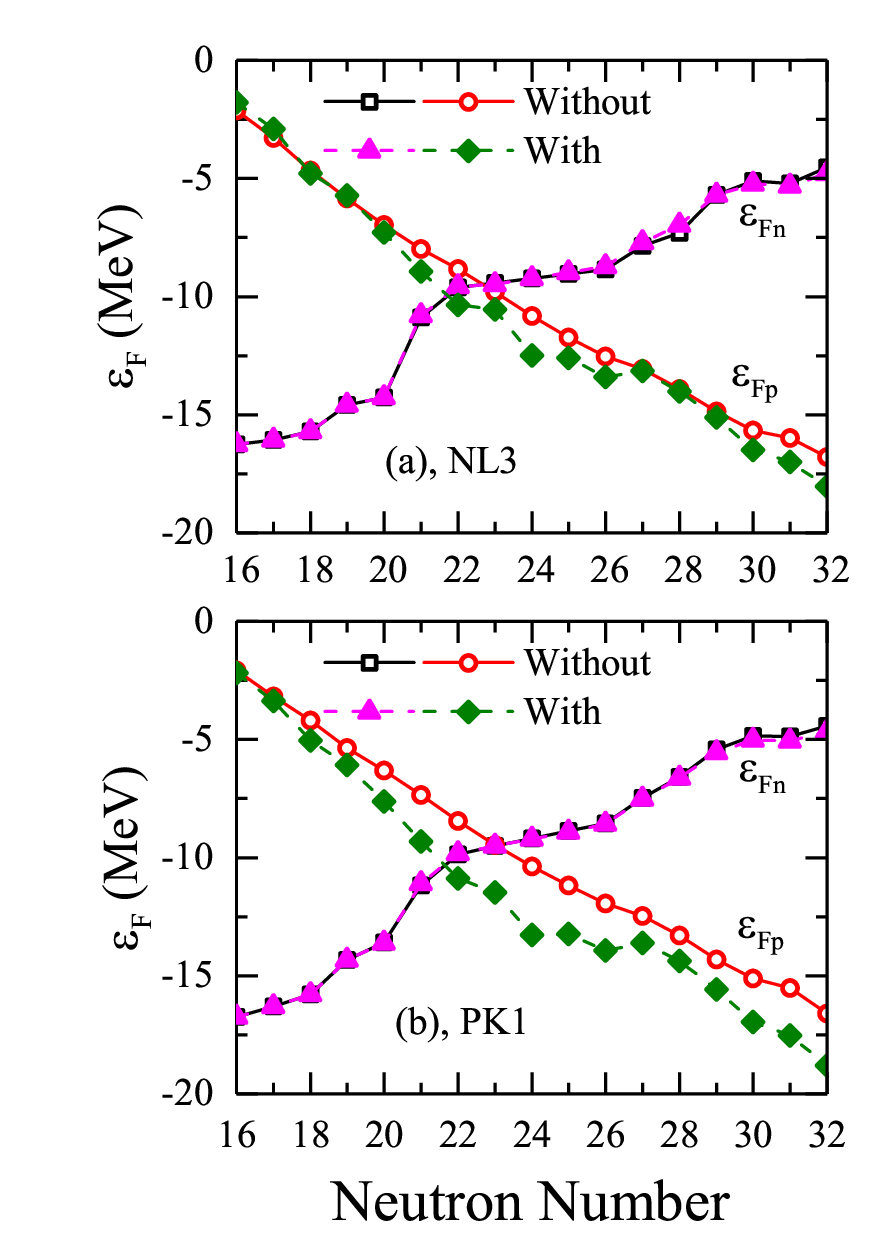}
\caption{(Color online) Fermi energies of neutrons ($\varepsilon_{\mathrm{Fn}}$) and protons ($\varepsilon_{\mathrm{Fp}}$) of calcium isotopes with and without making constraint on the rms charge radius are depicted as a function of neutron numbers through effective forces NL3 (a) and PK1 (b).} \label{fig2}
\end{figure}

The range of proton and neutron matter distributions is related to the disparity of the Fermi energies between protons and neutrons in mean-field calculations~\cite{PhysRevC.69.024318}.
In addition, charge radii can be influenced by the nuclear Fermi surface as well~\cite{PhysRevC.53.1599,PhysRevC.107.054307}.
Therefore, it is instructive to present the neutron ($\varepsilon_{\mathrm{Fn}}$) and proton ($\varepsilon_{\mathrm{Fp}}$) Fermi energies along calcium isotopes in Fig.~\ref{fig2}.
It is mentioned that the neutron Fermi energies obtained by effective forces NL3 and PK1 are almost not changed with the radius constraint calculations.
For proton Fermi energies, the apparent divergence occurs between the fully filled $N=20$ and $N=28$ shells.
Beyond $N=29$, this divergence is also presence.
%The Fermi energy difference between neutrons and protons is associated with the nuclear symmetry energy~\cite{PhysRevC.87.034327,qiu2023bayesian}.
Meanwhile, it is notably mentioned that the trend of changes of charge radii is significantly influenced by the isospin-dependence interactions~\cite{PhysRevLett.74.3744,PhysRevC.94.024304}.
This seems to suggest that the abnormal change in the charge radii can be reflected from the associated isospin-dependence interactions.
\begin{figure*}[htbp]
\centering
\includegraphics[scale=0.5]{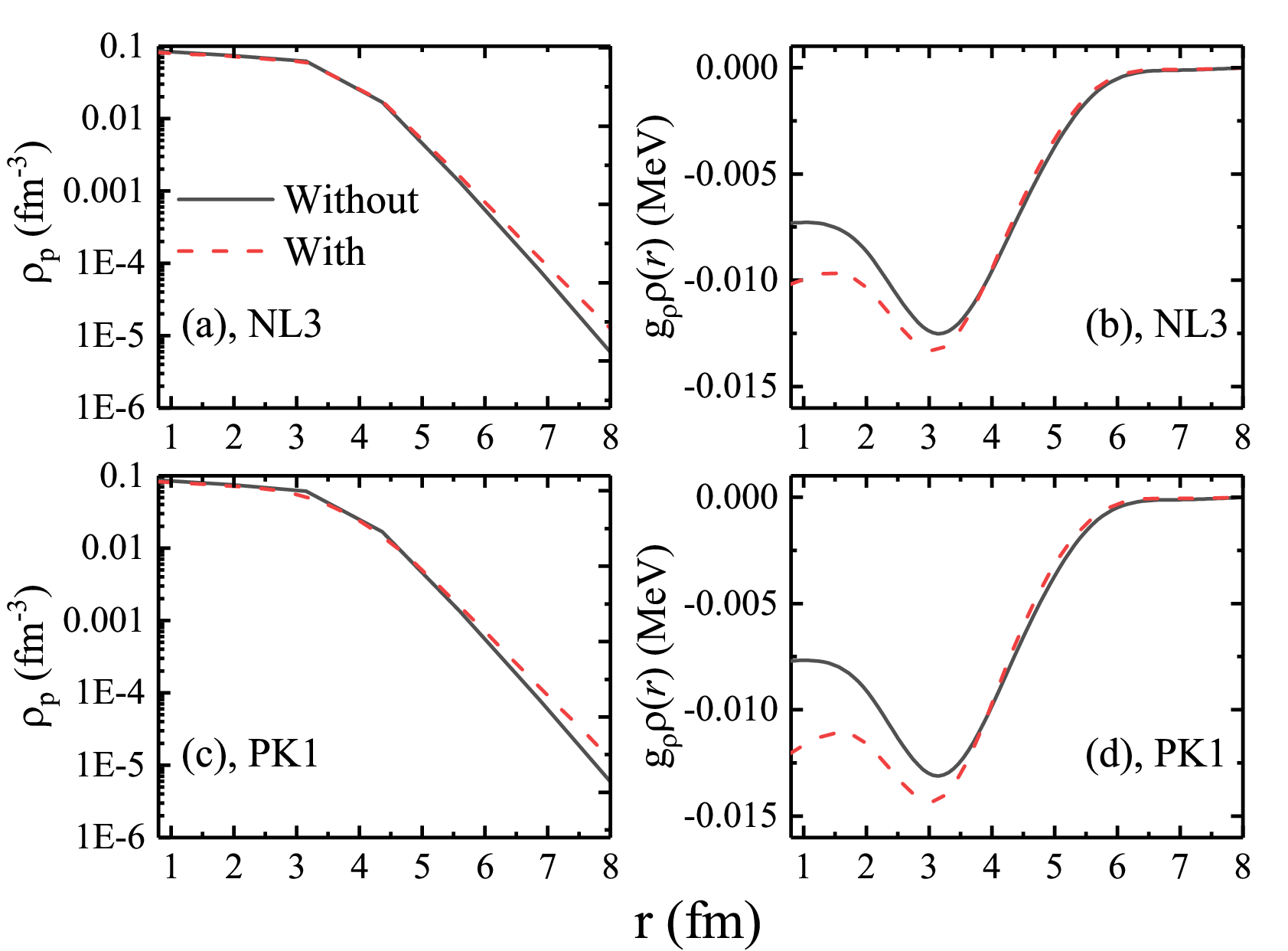}
\caption{(Color online) Proton density $\rho_{\mathrm{p}}$ and equivalent $\rho$ meson field $\mathrm{g}_{\rho}\rho(\bm{r})$ in $^{44}$Ca are depicted by the effective forces NL3 (a, b) and PK1 (c, d).} \label{fig30}
\end{figure*}

In Fig.~\ref{fig30} (a) and (c), the proton density distributions $\rho_{\mathrm{p}}(\bm{r})$ of $^{44}$Ca are shown with effective forces NL3 and PK1.
The constrained results give the relatively larger density distributions in the outer edge ($r>6$ fm).
This suggests that the corresponding central density is reduced.
Shown in Ref.~\cite{Miller:2018mfb}, it points out that the short-range neutron-proton correlation causes the increasing charge radius, namely exists more strong isospin-dependence interactions.
Under the relativistic mean field model, the $\rho$ meson is employed to describe the isospin asymmetry degrees.
In Fig.~\ref{fig30} (b) and (d), the equivalent meson field $\mathrm{g}_{\rho}\rho(\bm{r})$ in $^{44}$Ca are also presented.
It shows that the constrained calculations give more bound isospin-dependence interactions, namely rather strong isospin couplings.
As demonstrated in Ref.~\cite{PhysRevLett.74.3744}, the entire isospin dependence of the spin-orbit field is rather small in relativistic calculations.
%{\color{red}This may indicate that the charge radii of $^{42-46}$Ca isotopes cannot be described well in the relativistic mean field model.}
Therefore, the isospin-dependence components in effective nuclear potentials should be considered appropriately.

Actually, a strong coupling between neutron- and proton-pairing correlations leads to the anomalous behaviors in nuclear charge radii~\cite{ZAWISCHA1985309}.
Furthermore, recent studies suggest that $\alpha$-cluster structure has been performed in light or medium mass
nuclei~\cite{PhysRevC.100.034320,ADACHI2021136411,PhysRevLett.131.242501,PhysRevC.108.044314,Zhou2023}.
This means the more strong isospin-dependence interactions deriving from the neutron and proton pairs condensation should be taken into account in describing the fine structures of finite nuclei size.
Along calcium isotopic chain, the OES behavior in charge radii can be evidently observed~\cite{ANGELI201369,LI2021101440}.
As shown in Fig.~\ref{fig2}, these anomalous behaviors may be reflected from the changes of the proton Fermi surface.

With the increasing neutron numbers, the gradually deep proton single-particle levels are obtained along a long isotopic chain due to the strong isospin-dependence interactions~\cite{PhysRevC.61.047302,PhysRevC.81.051302}.
As encountered in proton Fermi energy, the constraint calculation makes the proton single-particle levels more bound.
For $^{44}$Ca, the constraining result obtained by NL3 force gives the proton single-particle energy $-14.54$ MeV for 2$s_{1/2}$ level and $-14.98$ MeV for $1d_{3/2}$ level. For results without making constraint, the proton single-particle energies are $-12.65$ MeV and $-13.48$ MeV for the corresponding 2$s_{1/2}$ and $1d_{3/2}$ levels, respectively. However, the neutron single-particle levels are almost unchanged before and after the constrained calculations. The same scenario can also be encountered in the PK1 parameter set, namely the constrained approach gives deeper proton single-particle levels.
Meanwhile, before and after making the charge radius constraint calculations, the proton and neutron occupation probabilities are almost unchanged for these NL3 and PK1 effective forces.
As shown in Fig.~\ref{fig2}, the proton Fermi energies are decreased with the increasing isospin asymmetry.
The constrained results reproduce the enlarged rms charge radius between $^{40}$Ca and $^{48}$Ca.
Meanwhile, the corresponding proton Fermi energies become lower in comparison to the cases without making constraint.
This may suggest that the isospin-dependence interactions in mean-field model cannot be captured adequately yet.

As shown in Table~\ref{tab1}, the binding energies of calcium isotopes are also changed by using the charge radius constraint approach. Large deviations of the proton Fermi energies can be observed apparently between the approaches with and without considering the constrained calculations. For $^{43}$Ca, the proton Fermi energy obtained by the charge radius constraint approach with PK1 set is $-11.47$ MeV, but the result obtained without constraining the charge radius is $-9.45$ MeV.
For NL3 set, before and after the constrained calculations, the corresponding proton Fermi energies are $-9.80$ MeV and $-10.53$ MeV, respectively. The same scenario can also be encountered in $^{44}$Ca where the constrained calculation gives the proton Fermi energy is $-13.27$ MeV for the PK1 set and $-12.48$ MeV for the NL3 set. By contrast, the results obtained without constraining the charge radius give the values of $-10.37$ MeV for the PK1 set and $-10.82$ MeV for NL3 set.

\begin{figure*}[htbp]
\centering
\includegraphics[scale=0.5]{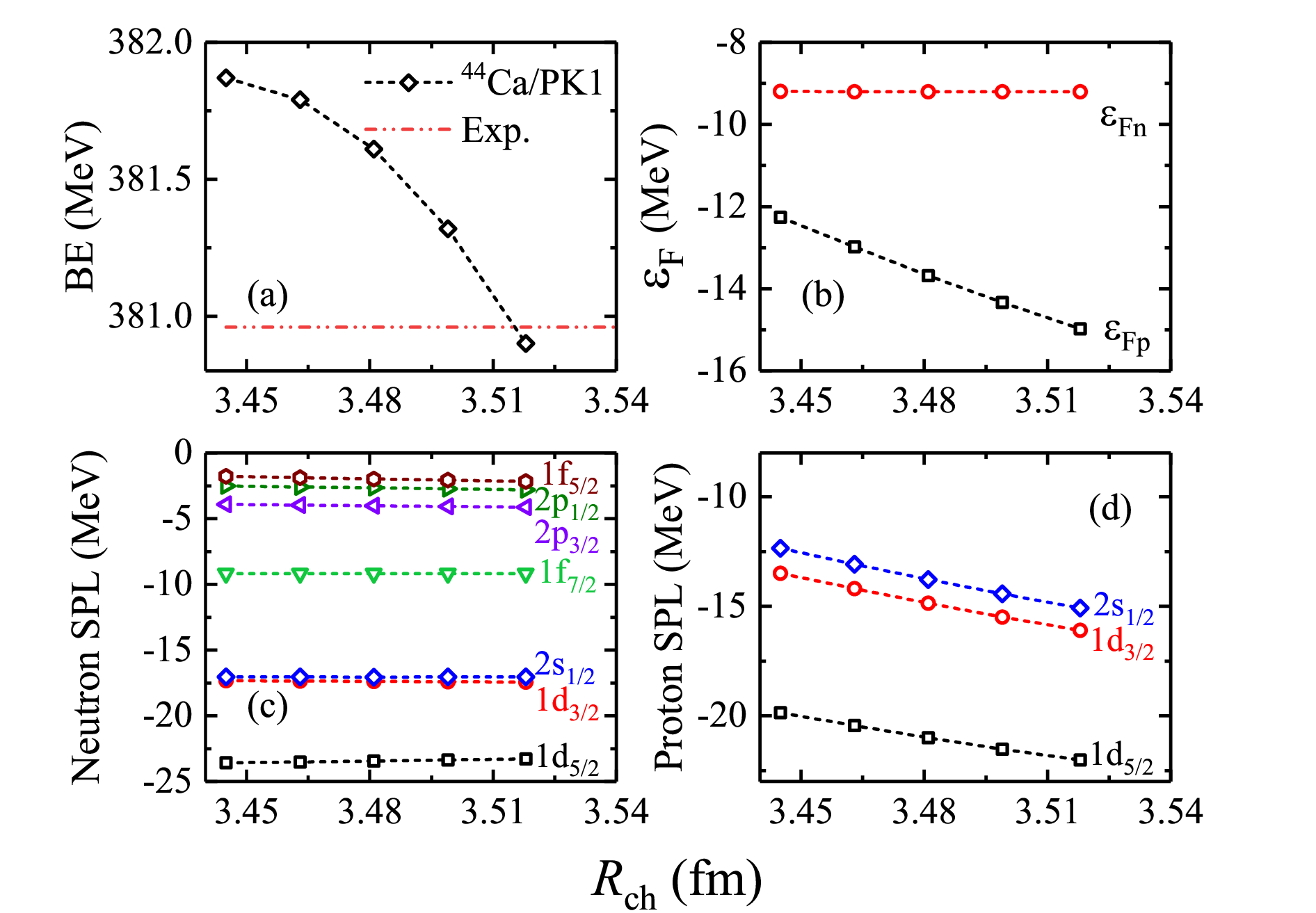}
\caption{(Color online) The binding energies (a), Fermi energies $\varepsilon_{\mathrm{F}}$ (b), neutron single-particle levels (c), and proton single-particle levels (d) of $^{44}$Ca isotope are depicted as a function of the constrained charge radius with effective force PK1. The experimental binding energy of $^{44}$Ca isotope is taken from Ref.~\cite{cpc2021}.} \label{fig3}
\end{figure*}
Before and after the constrained calculations, the apparent differences can be found in the binding energies, charge radii, proton density distributions, and the proton Fermi surface, etc. Furthermore, it is necessary to analyze in detail the changes of the binding energy, neutron and proton Fermi surface, shape deformation, neutron and proton single-particle levels, and the corresponding occupation probabilities with respect to the constrained charge radii. As shown in Fig.~\ref{fig3}, the binding energies, Fermi energies $\varepsilon_{\mathrm{F}}$, neutron single-particle levels, and proton single-particle levels of $^{44}$Ca isotope are depicted as a function of the constrained charge radius with effective force PK1. From this figure, one can find that the binding energy of $^{44}$Ca is smoothly decreased with the increasing values of the constrained charge radii. Meanwhile, as shown in Fig.~\ref{fig3} (b) and (c), the neutron Fermi energy $\varepsilon_{\mathrm{F}n}$ and neutron single-particle levels (SPLs) are almost not changed with the increasing values of the constrained charge radii. By contrast, as shown in Fig.~\ref{fig3} (b) and (d), the proton Fermi energy $\varepsilon_{\mathrm{F}p}$ and proton SPLs are almost linearly decreased with respect to the target charge radius of the constrained calculations. With the increasing charge radii, the corresponding occupation probabilities are almost not changed. In addition, the absolute values of the deformation parameters $\beta_{20}$ are less than 0.0005. This means the nucleus $^{44}$Ca is almost spherical shape with the increasing charge radii. That is why the spherical quantum numbers are used to remark the evolution of the neutron and proton single-particle levels in Fig.~\ref{fig3}, respectively.

\begin{figure}[htbp]
\centering
\includegraphics[scale=0.5]{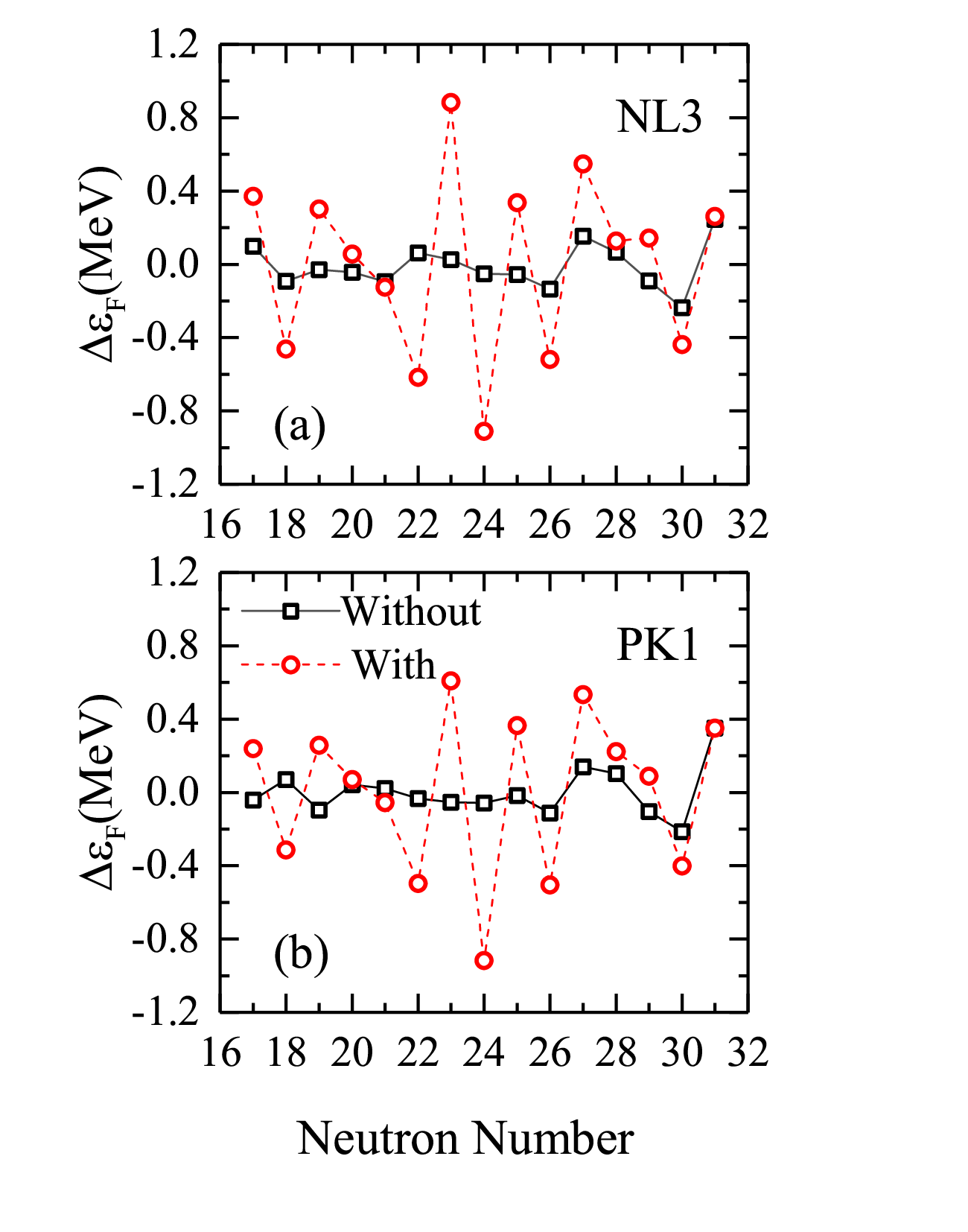}
\caption{(Color online) The values of $\Delta\varepsilon_{\mathrm{F}}$ derived from the proton Fermi energies against calcium isotopes as a function of neutron numbers with effective forces NL3 (a) and PK1 (b).} \label{fig4}
\end{figure}
In order to inspect these local variations of the proton Fermi energy, Eq.~(\ref{oef}) is still employed to facilitate these discontinuous behaviors.
In Fig.~\ref{fig4}, the values of $\Delta\varepsilon_{\mathrm{F}}$ derived from the proton Fermi energies of $^{36-52}$Ca isotopes are depicted with NL3 and PK1 forces, respectively.
The OES pattern of the proton Fermi energies is significantly observed along the calcium isotopes after the charge radius constraint is imposed.
As shown in Fig.~\ref{fig4}, we can find that the proton Fermi energies represent the inverse odd-even oscillations in comparison with the corresponding binding energies and charge radii. %The same scenarios can also be found in the proton Fermi energies with the increasing neutron numbers.
It should be mentioned that the OES behaviors of the proton Fermi energies are weakened at the neutron numbers $N=20$ and $28$ under the NL3 effective force. The same scenario can be encountered in the PK1 set as well.
The emergence of neutron magic numbers can be revealed through the systematic evolution of nuclear charge radii along a long isotopic family~\cite{PhysRevC.102.024307,An:2023kni}.
As demonstrated in Ref.~\cite{PhysRevC.105.014325}, the weakened OES behaviors can be observed in nuclear charge radii due to the strong shell closure effect.
This may result from the weakened OES of the proton Fermi energies at the completely filled shells.
The OES behaviors of the proton Fermi energies may provide an alternative access to understand the anomalous behaviors in nuclear charge radii. Therefore, the further investigation should be performed in the proceeding study.

\section{Summary and Outlook}\label{sec3}
Complementary to the anomalous behavior of the charge radii along calcium isotopes, the constraint calculations on root-mean-square (rms) charge radii are performed. The calculated results suggest that the amplitudes of odd-even staggering in the binding energies and two-neutron separation energies are almost similar with these two approaches, and agree with the experimental data well.
The weakening OES of charge radii can be regarded as local irregularity along calcium isotopic chain at the neutron numbers $N=20$ and $28$. This provides a signature to identify the shell closure effect~\cite{PhysRevC.100.044310,PhysRevC.105.014325}.
However, the empirical energy gap shows the larger amplitudes in comparison with the adjacent counterparts.

The rather strong isospin-dependence interactions can be deduced from the reproduced charge radii.
In order to review the OES behaviors in charge radii, the local variations of the proton Fermi energies for calcium isotopes are delineated by three-point formula~Eq.~(\ref{oef}).
It is obviously mentioned that the OES in the proton Fermi energies show the inverse amplitudes with respect to binding energies and charge radii.
Meanwhile, the weakening OES behaviors can also be found in the Fermi energies at $N=20$ and $28$.
As demonstrated in Refs.~\cite{PhysRevC.69.024318,PhysRevC.107.054307}, the nuclear size can be affected by the Fermi surface or the single-particle levels near the Fermi surface.
Hence this may offer an explanation why the enhanced odd-even oscillations of charge radii can be observed between $^{40}$Ca and $^{48}$Ca.

The calculated results suggest that the suitable isospin-dependence interactions should be supplied in relativistic mean field model.
Actually, nuclear charge radii are influenced by various mechanisms, such as high order moment~\cite{PhysRevC.101.021301,PhysRevC.104.024316}, isospin symmetry breaking~\cite{PhysRevC.106.L061306,PhysRevC.107.064302}, quadrupole deformation~\cite{An_2022}, etc.
Precision determination of charge radii plays a crucial role in nuclear physics.
The linear correlations between the charge radii differences of mirror-pair nuclei and the slope parameter of symmetry energy have been built to ascertain the isospin components of asymmetric nuclear matter interactions.
As shown in Refs.~\cite{PhysRevLett.130.032501,GAIDAROV2020122061,PhysRevC.97.014314,an2024}, the proton radii of mirror-pair nuclei can provide the opportunity to encode the information about neutron skin thickness.
This suggests that neutron skin thickness and charge radii are mutually determined.
Thus, this challenges us to perform high-precision descriptions in the charge radii of exotic nuclei.
\section{Acknowledgements}
This work was supported by the Natural Science Foundation of Ningxia Province, China (No. 2024AAC03015), the Open Project of Guangxi Key Laboratory of Nuclear Physics and Nuclear Technology, No. NLK2023-05, the Central Government Guidance Funds for Local Scientific and Technological Development, China (No. Guike ZY22096024), and the Key Laboratory of Beam Technology of Ministry of Education, China (No. BEAM2024G04). X. J. was grateful for the support of the National Natural Science Foundation of China under Grants No. 11705118, No. 12175151, and the Major Project of the GuangDong Basic and Applied Basic Research Foundation (2021B0301030006). N. T. was grateful for the support of the key research and development project of Ningxia (Grants No. 2024BEH04090) and the Key Laboratory of Beam Technology of Ministry of Education, China (No. BEAM2024G05). L.-G. C. was grateful for the support of the National Natural Science Foundation of China under Grants No. 12275025, No. 11975096 and the Fundamental Research Funds for the Central Universities (2020NTST06). F.-S. Z. was supported by the National Key R$\&$D Program of China under Grant No. 2023YFA1606401 and the National Natural Science Foundation of China under Grants No. 12047513, No. 12135004, No. 11635003, No. 11961141004, No. 11025524, No. 11161130520, the National Basic Research Program of China under Grant No. 2010CB832903.

\vspace*{2mm}
%\begin{small}\baselineskip=10pt\itemsep-2pt
%\bibliographystyle{CommTP}
%\bibliography{shn}

%\begin{small}\baselineskip=10pt\itemsep-2pt

\end{document}